# Frequency-Selective Rain Attenuation on Terahertz Channels

Yuheng Song[1], Wanzhu Chang[1], Kefeng Huang[1], Kaixin Sun[1], Weidong Hu[1,2], Jianjun Ma[1,2]

[1]School of Integrated Circuits and Electronics, Beijing Institute of Technology, Beijing 100081, China.

[2]State Key Laboratory of Environment Characteristics and Effects for Near-space, Beijing 100081, China.

Corresponding author: Jianjun Ma

## Abstract

Rain introduces broadband and frequency-selective attenuation in wideband terahertz (THz) links, making it necessary to identify a compact spectral descriptor that captures how the dominant loss region evolves with rainfall conditions. This article investigates the peak-frequency behavior of rain attenuation by combining Mie theory calculations with one separable laboratory Gaussian drop-size distribution (DSD) and eight outdoor empirical DSD models whose spectral shapes vary with rainfall rate. The analysis compares total loss, absorption, and scattering components, examines the roles of the characteristic DSD scale and representative drop size statistics, and evaluates the effect of temperature on the peak location. The results show that, unlike the fixed-shape laboratory case where the peak frequency remains unchanged with rainfall rate, all outdoor empirical DSD models exhibit a monotonic migration of the attenuation peak toward lower frequencies as rainfall rate increases. This migration exhibits quasi-exact inverse scaling with the rainfall-dependent DSD characteristic scale, is well described by a family-specific asymptotic power law in rainfall rate, and is governed mainly by characteristic drop size rather than by total drop concentration or fixed-temperature dielectric dispersion.

## 1 Introduction

The push toward sixth-generation (6G) wireless networks has elevated the terahertz (THz) band (0.1-10 THz) from a laboratory research topic to a serious candidate for ultra-high-capacity data links (Othman et al., 2025). Its suitability for high-capacity wireless communication scenarios has been investigated extensively (Federici et al., 2016; Ma, Song, et al., 2025; Rappaport et al., 2019). Studies surveying opportunities and challenges at frequencies above 100 GHz have noted that, although atmospheric gaseous absorption defines several discrete transmission windows in this regime, rain imposes a broadband, weather-dependent ceiling on link availability (Ma, Li, et al., 2025). Accurate characterization of rain-induced channel degradation is therefore essential for system design, band selection, and link-availability assessment in future high-frequency wireless systems. More critically, for wideband THz channels, the impact of rain cannot be fully captured by power loss evaluated at a few discrete carrier frequencies (Li et al., 2025; Li et al., 2023; Song et al., 2025). A spectrally resolved quantity is required to describe the frequency-selective structure of rain-induced degradation - one that conveys how the dominant loss region shifts across the band as rainfall conditions evolve.

Many experimental works have quantified rain-induced power loss across the THz band (see Table 1). A 10-Gbit/s outdoor channel measurement at 120 GHz over a 400 m path (Hirata et al., 2009) demonstrated that annual cumulative attenuation statistics agree well with ITU-R P.838-3 and Laws-Parsons model (with Mie theory) predictions. Simultaneous outdoor measurements at 77 GHz and 300 GHz (Norouzian et al., 2019) revealed that power loss at 300 GHz is comparable to, rather than substantially higher than, that at 77 GHz, a finding consistent with a post-peak plateau in the attenuation spectrum (Ma, Song, et al., 2025). The same study showed that Mie scattering calculations based on a Weibull drop-size distribution (DSD) achieve better agreement with measurements at 300 GHz than the Marshall-Palmer model. Controlled laboratory experiments have enabled more precise isolation of DSD effects. Using THz time-domain spectroscopy (THz-TDS) over 0.1-1 THz in a 4 m rain chamber, close agreement between measured attenuation and Mie-scattering predictions was demonstrated (Ma et al., 2015a, 2015b). Combined indoor and outdoor measurements at 140 GHz (Li et al., 2025) further established that, even at identical rainfall rates, different DSDs produce markedly different channel degradation. A population of smaller drops at higher number concentration induces stronger attenuation and larger power fluctuations than fewer, larger drops at the same rain rate. Outdoor measurements over a 54 m channel at 220-230 GHz (Song et al., 2025) subsequently confirmed that adopting a log-normal DSD in Mie calculations substantially improves agreement with measured data relative to the Marshall-Palmer model.

On the theoretical side, the existence of a broad rain-attenuation maximum below 1 THz has long been recognized. Power-law $aR^b$ coefficients were computed from Mie theory over 1-1000 GHz for several DSDs and temperatures, with the attenuation maximum already apparent in their tabulated coefficients (Olsen et al., 1978), and Mie calculations at millimeter and submillimeter wavelengths under Weibull and log-normal DSDs display the same feature (Sekine et al., 1987; Utsunomiya & Sekine, 2005). Calculations covering up to 1 THz under the Marshall-Palmer, Best (Best, 1950), Polyakova-Shifrin (Litovinov, 1957), and Weibull models established that rain-induced power loss reaches a maximum near 100 GHz and remains approximately constant or decreases slightly at higher frequencies (Seishiro Ishii et al., 2016). A 12-year disdrometer dataset acquired in Madrid was used to characterize attenuation from 80 to 200 GHz (Riera et al., 2023), finding that ITU-R P.838-3 agrees reasonably well at rainfall rates below approximately 5 mm/h but underestimates power loss at higher rates, with DSD-induced variability becoming increasingly significant above 100 GHz. Analogous trends were confirmed by DSD-based Mie calculations benchmarked against channel measurements at 240, 270, and 300 GHz (Kim et al., 2023). From a system-performance perspective, accounting for the full probability density of rainfall rate rather than averaged statistics was shown to reveal an order-of-magnitude underestimation of symbol-error rate at THz frequencies (Kupferman & Arnon, 2022). The standard engineering model ITU-R P.838-3 expresses specific attenuation as $\gamma = kR^{\alpha}$, with frequency-dependent coefficients $k(f)$ and $\alpha(f)$ tabulated from 1 to 1000 GHz (ITU-R, 2003). Contrary to a purely monotonic picture, the published coefficients themselves imply an interior spectral maximum: $k(f)$ increases up to roughly 200 GHz ($k_H$=1.64 at 200 GHz) and then declines slowly (1.38 at 1000 GHz), while $\alpha(f)$ decreases toward a shallow minimum near 500 GHz. Thus the ITU-R formulation already encodes a rainfall-dependent peak frequency. Evaluating it from the published coefficients gives $f_{peak} \approx 231$, 190, and 149 GHz at $R$=1, 10, and 50 mm/h, respectively. What the recommendation does not provide is the dependence of this peak on DSD structure - its coefficients derive from Mie calculations under fixed representative drop spectra - nor a microphysical account of why and how the peak migrates (Seishiro Ishii et al., 2016; Kupferman & Arnon, 2022; Okamura & Oguchi, 2010).

Table 1. Research contributions on rain attenuation characterization in the terahertz bands

| Ref. | Freq. (GHz) | Rain rate (mm/h) | Investigation method | Contribution |
|---|---|---|---|---|
| (Liebe et al., 1991) | <1000 | N/A | Theoretical | Double-Debye permittivity model for water |
| (Mätzler & Martin, 2002) | 1-1000 | 0.1-100 | Theoretical | Decomposition of rain extinction into absorption/scattering |
| (ITU-R, 2003) | 1-1000 | N/A | Theoretical | Power-law $\gamma = kR^{\alpha}$ |
| (Ellison, 2007) | <25000 | N/A | Theoretical | 3-relaxation and 2-resonance permittivity model |
| (Hirata et al., 2009) | 57-134 | 10-80 | Outdoor measurement | 10-Gbit/s outdoor rain measurements |
| (Setijadi et al., 2009) | 1-1000 | 10-200 | Theoretical | Temperature and multiple-scattering effects on rain attenuation |
| (Okamura & Oguchi, 2010) | <1000 | 2.5-150 | Theoretical | Peak frequency at ~100 GHz explicitly noted |
| (Ma et al., 2015b) | 625 | <500 | Laboratory measurement | Controlled rain chamber for THz and IR channel performance comparison |
| (Seishiro Ishii et al., 2016) | 8-1000 | 1-100 | Theoretical | Mie calculations confirming ~100 GHz peak |
| (Rappaport et al., 2019) | 100-3000 | 0.25-100 | Theoretical | Rain flattens above 100 GHz |
| (Norouzian et al., 2019) | 77, 300 | <20 | Outdoor measurement | Simultaneous rain measurements |
| (Kupferman & Arnon, 2022) | 30-1000 | <150 | Theoretical | Rain-rate PDF effects on mm-wave/THz system performance |
| (Riera et al., 2023) | 80-200 | <100 | Outdoor measurement | 12-year DSD-based characterization |
| (Li et al., 2025) | 140 | 6.8-30.6<br>733-14296 | Indoor & outdoor measurement | DSD sensitivity demonstrated |
| (Kim et al., 2023) | 240, 270, 300 | <100 | Outdoor measurement | DSD-based and ITU-R prediction |
| (Song et al., 2025) | 220, 225, 229 | <25 | Outdoor measurement | Log-normal DSD improves fit |

The peak frequency of the rain-attenuation spectrum can serve as a physically meaningful and compact spectral descriptor. It encapsulates, in a single frequency-domain quantity, how the dominant loss region evolves with rainfall rate, DSD structure, and temperature, reflecting the combined influence of raindrop-size statistics and the competition between absorption and scattering. Although the existence and approximate location of the peak are well established (Seishiro Ishii et al., 2016; Okamura & Oguchi, 2010; Olsen et al., 1978; Utsunomiya & Sekine, 2005), the peak frequency has rarely been treated as an explicit spectral descriptor in its own right. Its quantitative dependence on the DSD characteristic scale, an empirical parameterization of its rainfall-rate migration across DSD families, and the contrast between fixed-shape and rainfall-dependent drop spectra has not been established. This work addresses that gap through a systematic Mie-theory-based investigation spanning nine DSD configurations - one laboratory Gaussian DSD and eight outdoor empirical models.

## 2 Laboratory Gaussian DSD Case

We adopt a multi-DSD modeling framework based on Mie-scattering theory rather than the empirical ITU-R P.838-3 model. This is because Mie-theory approaches combined with explicit DSD information have been shown to provide a more accurate description of measured rain-attenuation results in the sub-THz and THz bands (Ma et al., 2015a, 2015b). The ITU-R model does not distinguish among different rainfall types and their associated drop-size spectra (ITU-R, 2003; Riera et al., 2023), while DSD structure has been demonstrated to significantly influence attenuation characteristics (Li et al., 2025; Song et al., 2025).

We first revisit the laboratory measurements in the previous publication (Li et al., 2023), which reports both raindrop-size statistics and multifrequency rain-induced power loss. In the experiment, the effective THz channel path within the rainfall region was approximately 4 m, the rainfall rate ranged from roughly 50 to 500 mm/h, and attenuation was measured at 140, 220, 340, and 675 GHz. Raindrops were generated by a fixed nozzle device, yielding a mean droplet diameter of approximately $D$=1.9 mm and a diameter distribution predominantly confined to 0.5-3.5 mm. Because the nozzle geometry was fixed, the spectral shape of the DSD did not vary with rainfall rate. Differences across rainfall rates arose solely from changes in the total number concentration of raindrops. Under this constraint, the laboratory drop spectrum is reconstructed in a separable Gaussian form (Lee et al., 2004), as

$$N(D;R) = N_t(R)h(D) \qquad (1)$$

Here, $N(D;R)$ is the number density of raindrops of diameter D per unit volume, $N_t(R)$ is the total number concentration, defined as the integral of $N(D;R)$ over all diameters, and h(D) is the truncated normalized Gaussian shape function, with mean μ = 1.9 mm and standard deviation σ = 0.28 mm (corresponding to a variance of 0.08 $mm^2$), truncated to the observed diameter range Dmin = 0.5 mm to Dmax = 3.5 mm and carrying units of mm-1. By construction, all rainfall rates share the same h(D), and differences in the DSD are encoded entirely in the amplitude term $N_t(R)$.

The functional relationship between $N_t(R)$ and R follows from the standard definition of rainfall rate, which is the DSD-weighted integral of the terminal fall velocity $v(D)$ (Liao et al., 2020). Combining this definition with Eq. (1) yields $N_t(R) \propto R$ under the fixed-shape assumption (Pruppacher & Klett, 2012). Since raindrops in laboratory experiment were produced by the same nozzle and experienced the same free-fall process (Li et al., 2023), $v(D)$ is treated as identical across rainfall rates for a given diameter (Ma et al., 2015a, 2015b). The fixed shape of $h(D)$ then gives a linear relation as $N_t(R) = aR$ (Bringi et al., 2003; Ma et al., 2015a), where a is a frequency-independent proportionality constant. A joint multifrequency least-squares fit to the measured attenuation data at all four frequencies yields a = 97.26 $m^{-3}(mm/h)^{-1}$, providing a single value that minimizes overall residuals and reduces sensitivity to frequency-specific measurement uncertainty. Although a is calibrated on the 50-500 mm/h laboratory range and subsequently applied at 1-50 mm/h, this extrapolation is exact by construction under the fixed-shape assumption, since Nt enters only as a linear amplitude factor. The laboratory DSD is used here as a fixed-shape reference spectrum rather than as a model of natural rain. This fitted value, substituted into Eq. (1), fully specifies the Gaussian DSD at each rainfall rate. Rain-induced power loss is then computed from the Mie-scattering integral (Zahid & Salous, 2022), as

$$\gamma_j\left(f,R,T\right) = 4.343\times10^3 \int_{D_{\min}}^{D_{\max}} N\left(D;R\right)\sigma_j\left(D,f,T\right)dD \quad \left(\text{dB/ km}\right) \qquad (2)$$

where $\sigma_j(D, f, T)$ denotes the extinction, scattering, or absorption cross-section in $m^2$ (j = ext, sca or abs) of a raindrop of diameter $D$ (mm) at frequency $f$ and temperature $T$, and N carries units of $m^{-3}mm^{-1}$. All single-particle quantities in this work are computed from Mie theory for homogeneous water spheres. The complex refractive index of liquid water $m(f, T)$ is obtained from the double-Debye permittivity model (Liebe et al., 1991), applicable below 1 THz. At T=27 $^{o}$C, m varies from 4.75+$j$2.67 at 50 GHz to 2.12+$j$0.54 at 1 THz. The Mie series is truncated according to the convergence criterion of Wiscombe (Wiscombe, 1980). Cross-sections are evaluated on a diameter grid of 0.01 mm and a frequency grid of 1 GHz spanning 10-1000 GHz. Peak frequencies are located within the 50-1000 GHz band and refined by three-point parabolic interpolation, giving sub-GHz effective resolution. Two simplifications are retained throughout - raindrops are treated as spheres, although natural drops larger than approximately 1 mm are oblate (Szakall et al., 2010), and single scattering is assumed (Liebe et al., 1991). Both approximations primarily affect absolute attenuation levels and polarization behavior rather than the location of the spectral peak, which is the quantity of interest here.

Using the 4 m channel path length and the measured ambient temperature of approximately 17$^{o}$C, the estimated attenuation is shown in Fig. 1(a). The reconstructed Gaussian DSD model yields substantially closer agreement with the measured data at all four frequencies than the log-normal DSD (Li et al., 2023), confirming both the improved estimation capability and the validity of the separable formulation.

The total power loss spectra in Fig. 1(b) are evaluated at a unified temperature of $T$= 27°C, with rainfall rate $R$ ranging from 1 to 50 mm/h. The spectra are computed up to 1 THz, so that the spectral maximum is an interior extremum of the computed band rather than an artifact of the search boundary. Atmospheric gaseous absorption is excluded throughout (Best, 1950; S Ishii, 2010), so that the spectral characteristics of rain-induced loss can be examined in isolation. The resulting spectra exhibit a clear, stable single-peak shape with the peak frequency fixed at approximately 56 GHz across the entire rainfall-rate range. As $R$ increases, the rain-induced power loss grows in magnitude, but the spectral shape and peak location remain unchanged. The same invariance holds when absorption and scattering are examined separately. This behavior follows directly from the model structure in Eq. (1). Because the Gaussian DSD is the product of a rainfall-rate-dependent amplitude Nt (R) and a fixed normalized shape function $h(D)$, varying R scales the attenuation spectrum uniformly without modifying its spectral structure. The peak frequency is therefore entirely determined by $h(D)$ and the dielectric response of water, not by the total drop concentration.

It should be noted that the 56 GHz peak frequency obtained here does not contradict the broad peak near 100 GHz reported in previous studies (Seishiro Ishii et al., 2016; Okamura & Oguchi, 2010). The peak location is governed by the statistical characteristics of the DSD - in particular, its characteristic droplet-size scale - and is therefore DSD-dependent rather than universal. This dependence is examined in detail for outdoor empirical DSD models in the following sections.

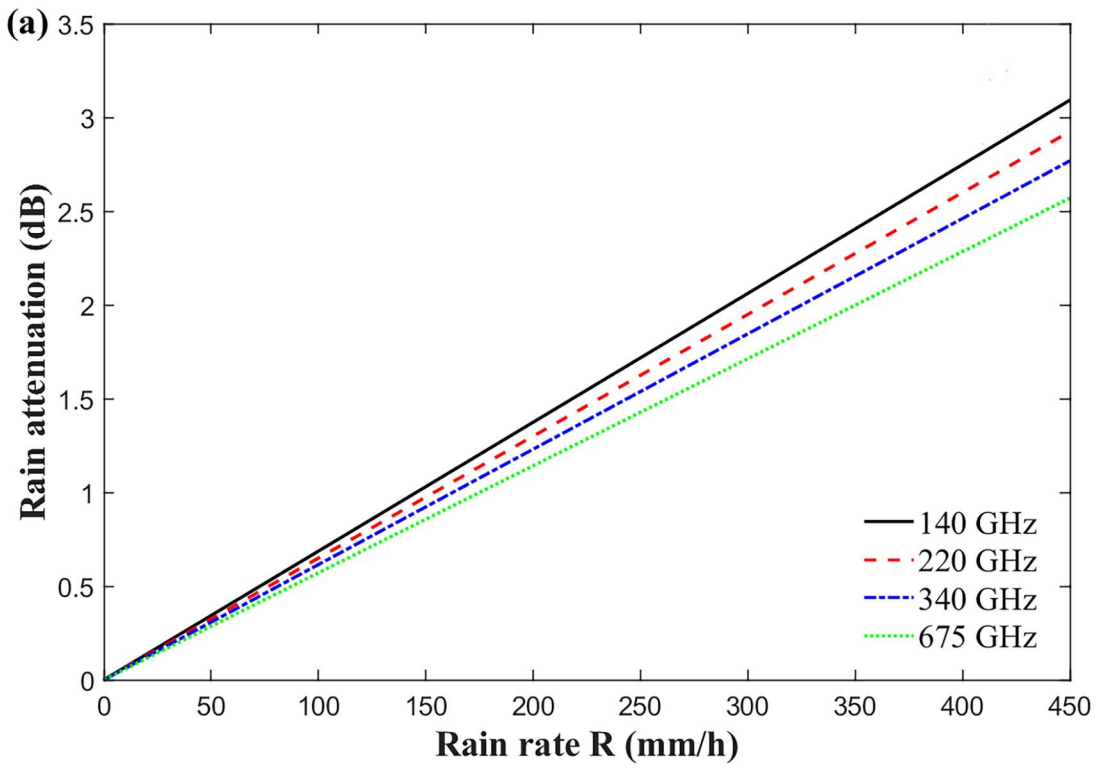

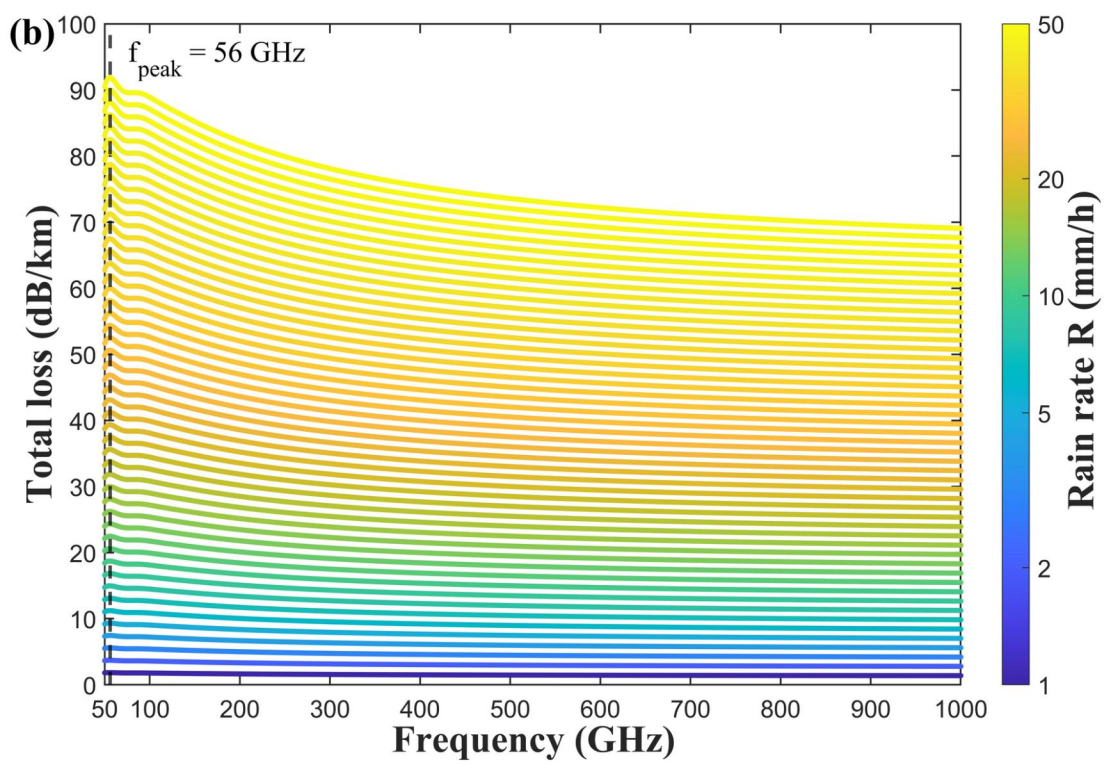

Figure 1. Rain-attenuation characteristics under the reconstructed indoor Gaussian DSD model: (a) rain attenuation versus rainfall rate for a 4 m path; (b) total-loss rain-attenuation spectra without atmospheric attenuation, with the corresponding peak frequency $f_{\text{peak}}$ indicated.

## 3 Outdoor Empirical DSD Models

Under natural outdoor rainfall, the DSD varies substantially with precipitation type and intensity (Long & Ulaby, 2015), in contrast to the fixed-shape laboratory condition analyzed in Section 2. To investigate the peak-frequency behavior across this broader range of microphysical conditions, several representative empirical DSD models are examined within a unified Mie-scattering framework. Long-term, strictly synchronized raindrop-spectrum observations of sufficient statistical stability are difficult to obtain in practice. A unified analysis based on multiple representative empirical models can therefore offer broader coverage of the natural-rain microphysical structures relevant to the peak-frequency problem, and enable general migration laws to be extracted from both commonalities and differences across models.

Eight DSD models drawn from three families are considered. The exponential family comprises the Marshall-Palmer (M-P) (Marshall & Palmer, 1948), Joss-Widespread (J-W) (Joss & Waldvogel, 1969), Joss-Drizzle (J-D) (Joss & Waldvogel, 1969), and Joss-Thunderstorm (J-T) (Joss & Waldvogel, 1969). The Atlas-Ulbrich Gamma family comprises the Hail, Sleet, and Snow models (Nemarich et al., 1988). The third family consists of the Best model (Best, 1950). Their mathematical forms are, respectively,

$$N(D;R) = N_0 \exp\left[-\Lambda(R)D\right] \qquad (3)$$

$$N(D;R) = N_0(R)D^2 \exp\left[-\Lambda(R)D\right] \qquad (4)$$

$$N(D;R) = \frac{13.5W(R)}{\pi a^4(R)}\left(\frac{D}{a(R)}\right)^{-1.75} \exp\left[-\left(\frac{D}{a(R)}\right)^{2.25}\right] \qquad (5)$$

Equations (3), (4), and (5) correspond to the exponential (Song et al., 2025), Atlas-Ulbrich Gamma (Nemarich et al., 1988), and Best (Best, 1950) families, respectively. The model parameters, their units, and the precipitation regimes the parameterizations represent are listed in Table 2. All outdoor empirical DSD models are uniformly defined over the diameter range 0.1-8 mm (Szakall et al., 2010), which is more appropriate for representing the full natural-rain spectrum - from small to large drops - than the narrower range adopted in the indoor reference case.

Table 2. Parameter settings, units, and representative precipitation regimes for the empirical DSD models examined in this work (Seishiro Ishii et al., 2016; Song et al., 2025).

| Family | Model | Parameter relations | Parameter units |
| --- | --- | --- | --- |
| Exponential | M-P | $N_0 = 8000$, $\Lambda = 4.1R^{-0.21}$ | $N_0$ [m$^{-3}$ mm$^{-1}$]<br>$\Lambda$ [mm$^{-1}$] |
| | J-W | $N_0 = 7000$, $\Lambda = 4.1R^{-0.21}$ | |
| | J-D | $N_0 = 30000$, $\Lambda = 5.7R^{-0.21}$ | |
| | J-T | $N_0 = 1400$, $\Lambda = 3.0R^{-0.21}$ | |
| *Gamma* | Hail | $N_0 = 64500\mathrm{R}^{-0.5}$, $\Lambda = 6.95R^{-0.27}$ | $N_0$ [m$^{-3}$ mm$^{-3}$]<br>$\Lambda$ [mm$^{-1}$] |
| | Sleet | $N_0 = 11750\mathrm{R}^{-0.29}$, $\Lambda = 4.87R^{-0.20}$ | |
| | Snow | $N_0 = 2820\mathrm{R}^{-0.18}$, $\Lambda = 4.01R^{-0.19}$ | |
| Best | Best | $W = 67\mathrm{R}^{0.846}$, $a = 1.3R^{0.232}$ | $W$[m$^{-3}$ mm$^{3}$]<br>$a$ [mm] |

Because the eight parameterizations were fitted to different precipitation regimes, the rainfall-rate $R$ is not exactly equivalent across models. As a consistency check, the rainfall rate implied by each DSD is computed from the flux integral

$$R_{\mathrm{imp}} = 6\pi \times 10^{-4} \int_{D_{\min}}^{D_{\max}} D^3 v(D) N(D;R)\, dD \qquad (6)$$

with the terminal-velocity relation $v(D)$=9.65-10.3$e^{(-0.6D)}$ m/s (Atlas & Ulbrich, 1977) ($D$ in mm, $N$ in m$^{-3}$ mm$^{-1}$, $R_{\mathrm{imp}}$ in mm/h). For the exponential and Best models, $R_{\mathrm{imp}}$ tracks the nominal label to within roughly ±20% over 1-50 mm/h (M-P: 1.18, 11.6, and 54.6 mm/h at $R$ = 1, 10, and 50 mm/h; Best: 1.02, 10.2, and 47.4 mm/h), and the Gamma-Sleet and Gamma-Snow models behave similarly. The Gamma-Hail model deviates most strongly (0.44$R$ at 1 mm/h rising to 1.43$R$ at 50 mm/h), consistent with its frozen-phase origin (Nemarich et al., 1988). Cross-model comparisons at fixed nominal $R$ therefore compare slightly different water fluxes, and the resulting spread should be read as a plausible envelope of natural microphysical variability rather than as predictions for a strictly identical rain condition.

Under the unified temperature condition $T$ = 27$^{\circ}$C, the total-loss spectra of the Best DSD model are shown in Fig. 2(a). The spectrum preserves a clear single-peak structure across the entire rainfall-rate range considered. As $R$ increases, the total loss rises over the full frequency band, but the spectral evolution is not a simple amplitude scaling. The peak position shifts systematically toward lower frequencies with increasing rainfall rate. This constitutes a qualitative departure from the laboratory Gaussian case - the Best model spectra exhibit not only a magnitude increase but a genuine migration of the dominant loss region. Analogous behavior is observed for all other models considered, as shown in Fig. A1 in the Appendix. Under outdoor empirical DSD conditions, therefore, rainfall-rate variation alters the spectral structure itself rather than merely rescaling its magnitude.

The component-wise peak frequencies extracted from Fig. 2(b) further clarify the migration mechanism. The total-loss, absorption, and scattering peak frequencies all decrease monotonically with increasing $R$, but at evidently different rates. At low rainfall rates ($R < 5$ mm/h, with the precise threshold varying by DSD model), the scattering peak lies at the highest frequency and exhibits the most rapid downward shift as $R$ increases. The absorption peak migrates more gradually and becomes the highest-frequency component for $R > 5$ mm/h. The total-loss peak remains intermediate between the two throughout the full rainfall range. This ordering follows from the small-particle limits of the Mie efficiencies. Absorption grows as $Q_{\text{abs}} \propto \chi$ ($\chi$ is the standard Mie size parameter) while Rayleigh scattering grows as $Q_{\text{sca}} \propto \chi^4$, so the scattering contribution rises and saturates at larger size parameters - equivalently, higher frequencies for a given drop spectrum - than absorption. This interpretation is consistent with prior Mie-scattering studies showing that rain extinction arises from distinct absorption and scattering contributions and that the predicted attenuation depends strongly on the assumed DSD model (Li et al., 2025). Here, the component-wise peak-frequency trajectories show that the downward migration of the total-loss peak is associated with rainfall-dependent redistribution between absorption and scattering, rather than with either mechanism alone.

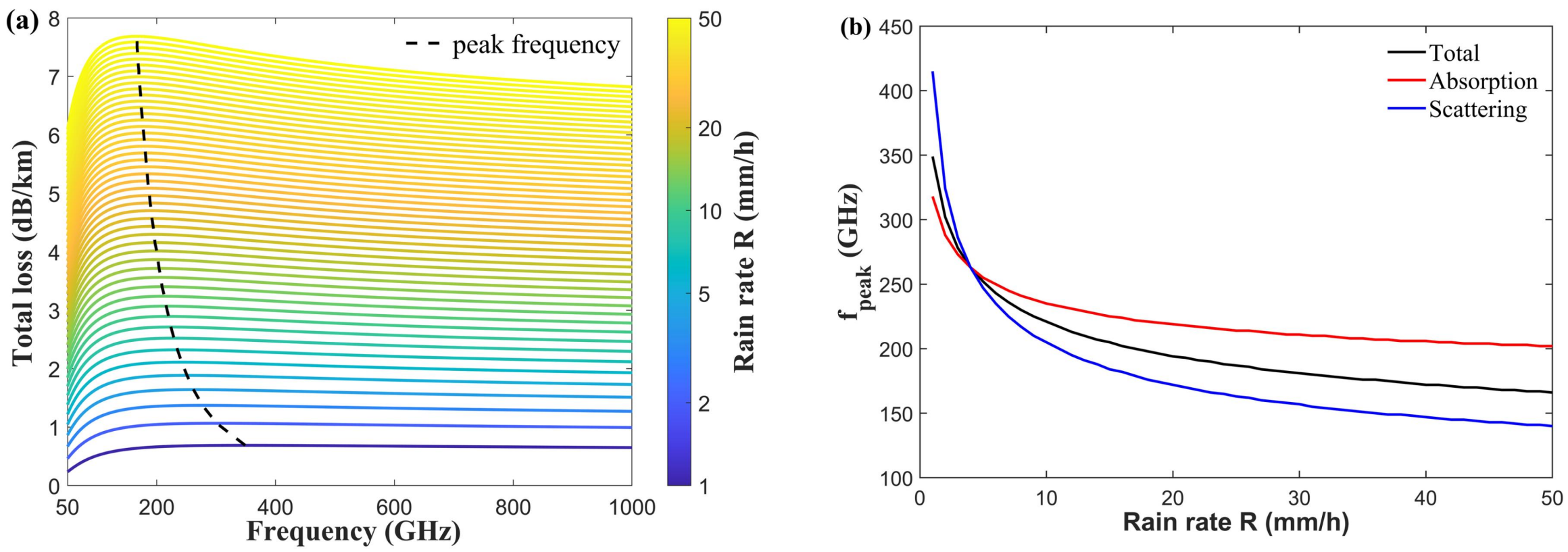

Figure 2. Rain-attenuation characteristics under the Best DSD model: (a) total-loss rain-attenuation spectra over the rainfall-rate range of 1–50 mm/h; (b) peak frequencies of the total-loss, absorption, and scattering components as functions of rainfall rate $R$.

To identify the physical quantities governing peak migration, the single-particle cross-section $\sigma_j(D,f,T)$ in Eq. (2) is expressed as, $\sigma_j(D,f,T) = Q_j(x,m(f,T))\cdot\pi(D/2)^2$ (Ma, Song, et al., 2025), where $m(f,T)$ is the complex refractive index of liquid water, $Q_j$ is the Mie efficiency factor with $j = ext$, $sca$ or $abs$. Following the normalization framework of (Lee et al., 2004), the empirical DSD is rewritten in the product form, as

$$N(D;R) = \frac{N_t(R)}{D_c(R)} g(u;R) \qquad (7)$$

where $u = D/D_c(R)$ is a dimensionless diameter variable, $g(u;R)$ is the dimensionless normalized DSD shape function, and $D_c(R)$ is the model-internal characteristic scale describing the rainfall-dependent stretching of the diameter axis. For the exponential and Gamma models, $D_c(R)=1/\Lambda(R)$; for the Best model, $D_c(R)=a(R)$. For all eight empirical models the shape function carries no explicit rainfall dependence - $g(u) \propto e^{-u}$ for the exponential family, $g(u) \propto u^2 e^{-u}$ for the Gamma family, and $g(u) \propto u^{-1.75}\exp(-u^{2.25})$ for the Best model - so all rainfall dependence enters through $N_t(R)$, $D_c(R)$, and the truncated integration limits. Substituting $D=D_c(R)u$ and $dD=D_c(R)du$ into Eq. (2) yields

$$\gamma_j(f,R,T) = K\frac{\pi}{4}N_t(R)D_c^2(R)\int_{D_{\min}/D_c(R)}^{D_{\max}/D_c(R)} g(u;R)u^2 Q_j\left(\frac{\pi D_c(R)fu}{c}, m(f,T)\right)du \quad (8)$$

where $K = 4.343\times10^3$. Because the peak condition $\partial\gamma_j/\partial f=0$ is implicit, $f_{\mathrm{peak},j}$ is not available in closed form. It is governed by three quantities apparent in Eq. (8): the DSD characteristic scale $D_c(R)$, the normalized shape function $g(u;R)$, and the dielectric response $m(f,T)$. This dependence is written schematically as $f_{\mathrm{peak},j}= F_j[D_c(R), g(u;R), m(f,T)]$, where $F_j$ denotes the peak-location mapping implied by Eq. (8) and the extremum condition, not an independent physical variable. Crucially, under fixed $R$ and $T$, the total number concentration $N_t(R)$ appears in Eq. (8) only as a multiplicative amplitude factor and therefore does not determine the peak position.

The remaining frequency dependence in Eq. (8) enters through two channels—the size-parameter argument $\pi D_c(R)fu/c$ and the explicit dispersion of $m(f,T)$. To isolate the latter, calculations using the full frequency-dependent refractive index $m(f,T)$ were compared at $T=$ 27°C with calculations in which the refractive index was fixed at $m(f_{\mathrm{ref}}, T)$, for $f_{\mathrm{ref}}$ = 140, 220, and 340 GHz, while all other settings were kept unchanged. The comparison covered the frequency range of 50–1000 GHz on a 1-GHz grid, all eight outdoor empirical DSD models, and the total-loss, absorption, and scattering components. In every tested case, the two treatments yielded the same extracted $f_{\mathrm{peak}}$ on the adopted frequency grid (see Fig. A2 in the Appendix). Thus, under the present fixed-temperature conditions, explicit dielectric dispersion produces no resolvable shift in the spectral peak, and the rainfall-dependent peak migration is governed primarily by the DSD characteristic scale $D_c(R)$. This conclusion concerns dispersion at fixed temperature and does not preclude the influence of temperature changes discussed in Section 5.

When the refractive index m is held fixed and the finite diameter limits are neglected for the purpose of scaling analysis, the attenuation spectrum depends on frequency f and the characteristic scale $D_c(R)$ only through the dimensionless variable $x = \pi f D_c(R)/c$. The peak condition therefore defines a model-family-dependent value $x^*$, giving $f_{\mathrm{peak}} = cx^*/[\pi D_c(R)]$. Exact inverse scaling would require $x^*$ to remain constant as rainfall rate changes. The computed peak positions closely follow this expectation: over R = 1–50 mm h$^{-1}$, $x^*$ remains within 0.84–0.95 for the exponential models, 0.40–0.47 for the Gamma models, and 4.75–5.60 for the Best model. Across the examined models, the full peak-to-peak variation in $x^*$, normalized by its mean within each model, is only about 12–16%, whereas $f_{\mathrm{peak}}$ itself changes by more than a factor of two. Thus, although $x^*$ is not strictly constant, its comparatively small variation shows that the rainfall-dependent change in $f_{\mathrm{peak}}$ is controlled predominantly by $D_c(R)$. The peak migration within each DSD model is therefore well described by a quasi-inverse relationship between $f_{\mathrm{peak}}$ and $D_c(R)$.

Within each empirical DSD family, the peak frequencies of total loss, absorption, and scattering are therefore all well approximated by a linear function of $1/D_c(R)$, as illustrated for the Best model in Fig. 3(a),

$$f_{\mathrm{peak},j}(R) \approx \frac{a_j}{D_c(R)} + b_j \quad (9)$$

where $a_j \approx (c/\pi)x_j^* = 95.5x_j^*$ GHz·mm, where $x_j^*$ denotes a representative value of the dimensionless peak parameter for DSD model $j$ over the fitted rainfall-rate range, and the offset $b_j$ accounting empirically for the small residual departure from exact inverse scaling. The fitted slope–offset pairs are broadly consistent with this prediction: for example, 75.1 GHz·mm with an offset of 12.6 GHz for the M–P model, compared with $(c/\pi)\times 0.87 \approx 83.2$ GHz·mm, and 35.5 GHz·mm with an offset of 15.6 GHz for the Gamma–Hail model, compared with $(c/\pi)\times 0.426 \approx 40.7$ GHz·mm. Equation (9) is therefore not merely an empirical observation but a theory-guided affine approximation to the inverse-scaling law. Since $D_c(R) \propto R^{p_{\mathrm{Dc}}}$, where $p_{\mathrm{Dc}} > 0$ is the characteristic-scale exponent from the parameterization in Table 2, Eq. (9) implies

$$f_{\mathrm{peak},j}(R) \approx c_j R^{-p_{Dc}} + b_j \qquad (10)$$

with $c_j$ determined by the loss component and the DSD model. Under fixed-temperature conditions, therefore, the peak frequency in outdoor empirical DSD models follows an approximately decaying power law in rainfall rate, with a nonzero constant offset.

The results in Fig. 2 exhibit two consistent features that a candidate empirical model must capture. $f_{\mathrm{peak}}$ decreases monotonically with $R$, and the decrease becomes progressively weaker at high rainfall rates, indicating an asymptotic flattening tendency. Motivated by Eq. (10), three candidate functional forms are evaluated. The asymptotic power-law (APL) form is (Newell & Rosenbloom, 2013)

$$f_{\mathrm{peak}}(R) = f_\infty + BR^{-p} \quad (11)$$

where $f_\infty$ is the asymptotic peak frequency, $B$ is an amplitude coefficient, and $p$ is the decay exponent. The term $BR^{-p}$ captures the dominant downward migration, while $f_\infty$ represents the high-rainfall plateau. To test whether an explicit transition parameter improves the fit, the Hill form (Motulsky & Christopoulos, 2004) is also considered, as

$$f_{\mathrm{peak}}(R) = f_\infty + \frac{B}{1+(R/R_c)^p} \qquad (12)$$

where $R_c$ is a characteristic rainfall rate governing the transition between the low-$R$ and high-$R$ regimes. To test whether the asymptotic offset is necessary at all, a simple power-law (PL) form (Newman, 2005) is included:

$$f_{\mathrm{peak}}(R) = BR^{-p} \qquad (13)$$

The three models (APL, Hill and PL) serve distinct diagnostic purposes. APL is the scale-motivated baseline, Hill tests the value of an explicit transition parameter, and PL tests whether the offset term can be omitted.

The fitted quantity in each case is the simulated total-loss $f_{\mathrm{peak}}$-$R$ relation at $T$=27°C for each empirical DSD model, extracted at $R$ = 1, 2, …, 50 mm/h (50 points) with the parabolic peak refinement of Section 2 and fitted by Levenberg-Marquardt nonlinear least squares. The APL and Hill forms are applied on $R \geq 1$ mm/h only, since the $BR^{-p}$ term diverges as $R \to 0+$. Fitting quality is evaluated using the root mean square error (RMSE), the coefficient of determination $R^2$ (denoted $R^2$ to avoid collision with the rainfall rate $R$), and the Akaike and Bayesian information criteria (AIC/BIC) (Akaike, 1974; Schwarz, 1978), which penalize model complexity. Results are summarized in Table 3. The APL model delivers the best overall performance across all eight DSDs. It achieves the smallest RMSE and the lowest AIC and BIC in every case, while maintaining $R^2$ consistently very close to unity. The Hill model yields comparable residuals, but its additional parameter is penalized by both criteria without a compensating improvement (for the Best model, AIC = -131.1 for APL versus -129.0 for Hill and +42.5 for PL). The PL model shows clearly larger deviations for most DSDs, confirming that a zero-offset decay law is insufficient to describe the peak-frequency evolution over the rainfall range considered.

Table 3. Fitting performance of the three candidate total-loss $f_{peak}$-$R$ models across all empirical DSD configurations at $T$=27°C. AIC = $n\ln(\mathrm{RMSE}^2)+2k$ and BIC = $n\ln(\mathrm{RMSE}^2)+k\ln n$, $n$ = 50 fitted points and k the number of free parameters (APL: $k$ = 3; Hill: $k$ = 4; PL: $k$ = 2).

| DSD | APL ($k$ = 3) | | Hill ($k$ = 4) | | PL ($k$ = 2) | |
|---|---|---|---|---|---|---|
| | RMSE | R² | RMSE | R² | RMSE | R² |
| Best | 0.2538 | 0.99995 | 0.2540 | 0.99995 | 1.4698 | 0.99843 |
| Gamma-Hail | 0.3528 | 0.99988 | 0.3536 | 0.99988 | 1.4161 | 0.99810 |
| Gamma-Sleet | 0.3067 | 0.99974 | 0.3081 | 0.99974 | 0.5498 | 0.99916 |
| Gamma-Snow | 0.2840 | 0.99965 | 0.2843 | 0.99965 | 0.3763 | 0.99939 |
| M-P | 0.3210 | 0.99992 | 0.3220 | 0.99992 | 1.3374 | 0.99859 |
| J-W | 0.3210 | 0.99992 | 0.3220 | 0.99992 | 1.3374 | 0.99859 |
| J-D | 0.3708 | 0.99995 | 0.3729 | 0.99995 | 2.3622 | 0.99808 |
| J-T | 0.2776 | 0.99988 | 0.2780 | 0.99988 | 0.9445 | 0.99856 |

The APL model is therefore adopted as the preferred form for subsequent analysis. The best-fit parameters are listed in Table 4. Although all eight models share the same APL structure, the fitted parameters ($f_\infty$, $B$, $p$) differ substantially across DSD families, reflecting differences in their characteristic droplet-size scales. Two caveats attach to Table 4. First, $f_\infty$ is the $R\to\infty$ asymptote of the fit, not a peak frequency realized within the 1-50 mm/h range; six of the eight fitted values lie below the 50-GHz lower edge of the peak-search band, underscoring its extrapolative character. Second, the fitted decay exponents $p$ closely track the characteristic-scale exponents $p_{Dc}$ of the underlying parameterizations (see Table 4), as Eq. (10) requires. The modest excess of $p$ over $p_{Dc}$ for some models reflects the same offset and drift terms discussed above.

Table 4. Best-fit APL parameters for each empirical DSD model at $T$=27°C. together with the characteristic-scale exponent $p_{Dc}$ of each parameterization

| DSD | Best model | $f_\infty$ | $B$ | $p$ | $p_{Dc}$ |
|---|---|---|---|---|---|
| Best | APL | 56.76 | 292.10 | 0.25 | 0.23 |
| Gamma-Hail | APL | 33.31 | 233.07 | 0.31 | 0.27 |
| Gamma-Sleet | APL | 23.69 | 163.81 | 0.21 | 0.20 |
| Gamma-Snow | APL | 15.71 | 141.26 | 0.19 | 0.19 |
| M-P | APL | 49.49 | 280.12 | 0.25 | 0.21 |
| J-W | APL | 49.49 | 280.12 | 0.25 | 0.21 |
| J-D | APL | 71.86 | 402.27 | 0.28 | 0.21 |
| J-T | APL | 39.90 | 201.95 | 0.24 | 0.21 |

The Best model serves as a representative example. As shown in Fig. 3(b), the APL curve reproduces the simulated $f_{peak}$-$R$ data over the full rainfall range with excellent agreement, capturing both the rapid decline at low $R$ and the gradual flattening at high $R$. Together with the linear dependence on $1/D_c$ established in Fig. 3(a), these results support the APL form as a compact and physically grounded empirical description of fixed-temperature peak-frequency migration. They further suggest that, for wideband THz channels, rain acts not only as additional loss but also as a spectral-reshaping mechanism, since THz channel impairments generally manifest through frequency-selective attenuation across the band (Rappaport et al., 2019).

A final qualification concerns the sharpness of the maximum. Defining the prominence width $W_5(R)$ as the frequency span over which the total loss remains within 5% of its maximum, the Best model gives $W_5$ = 728, 536, 460, 379, and 333 GHz at $R$ = 1, 5, 10, 25, and 50 mm/h, and the M-P model gives 523, 359, 307, 254, and 224 GHz. At low rainfall rates the spectrum is plateau-like over hundreds of GHz, and $f_{peak}$ is best interpreted as marking the onset of a broad high-loss plateau rather than a sharply localized maximum; the peak sharpens considerably as $R$ increases. For system-level use, $f_{peak}$ should accordingly be paired with a width metric of this kind.

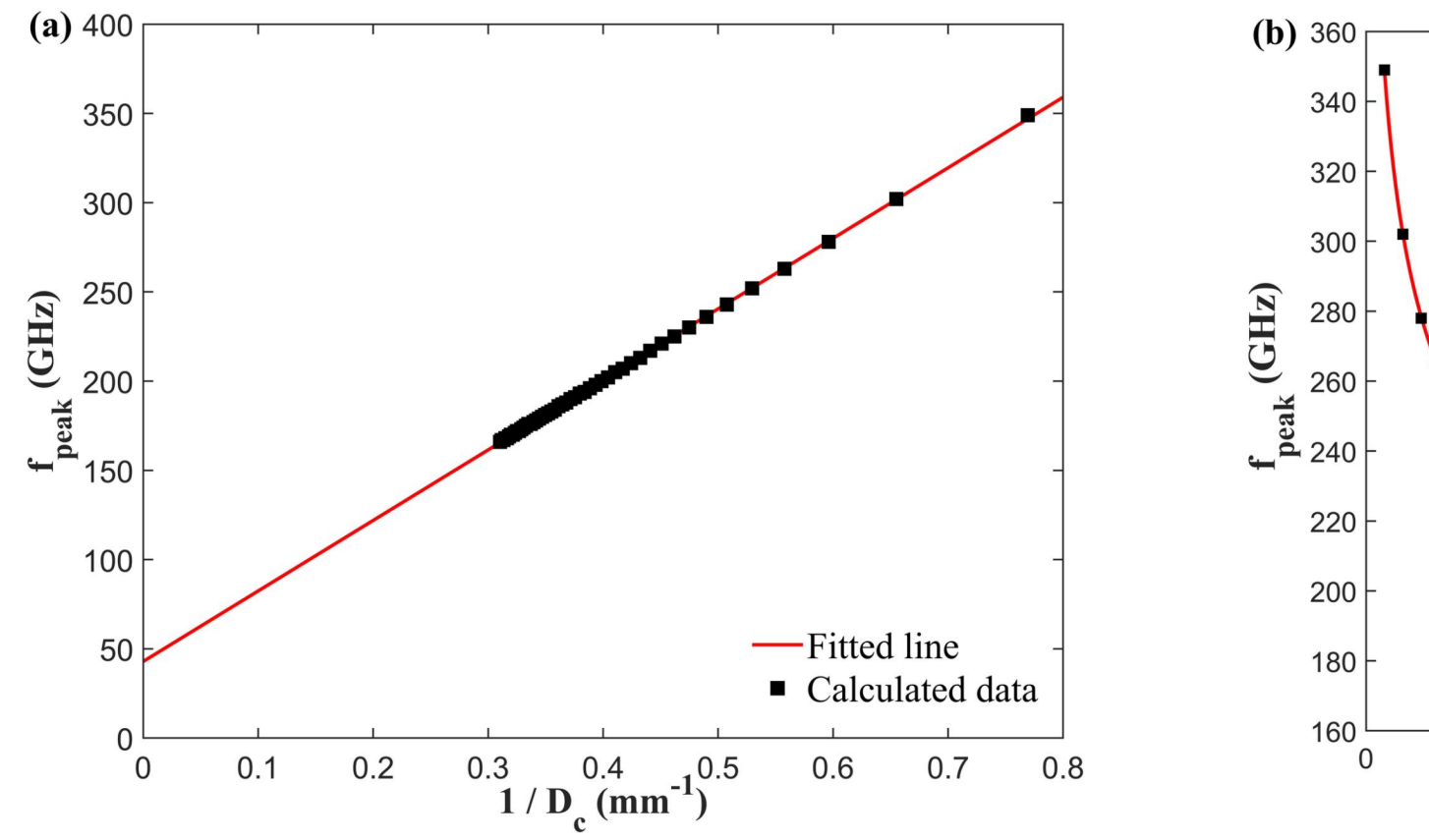

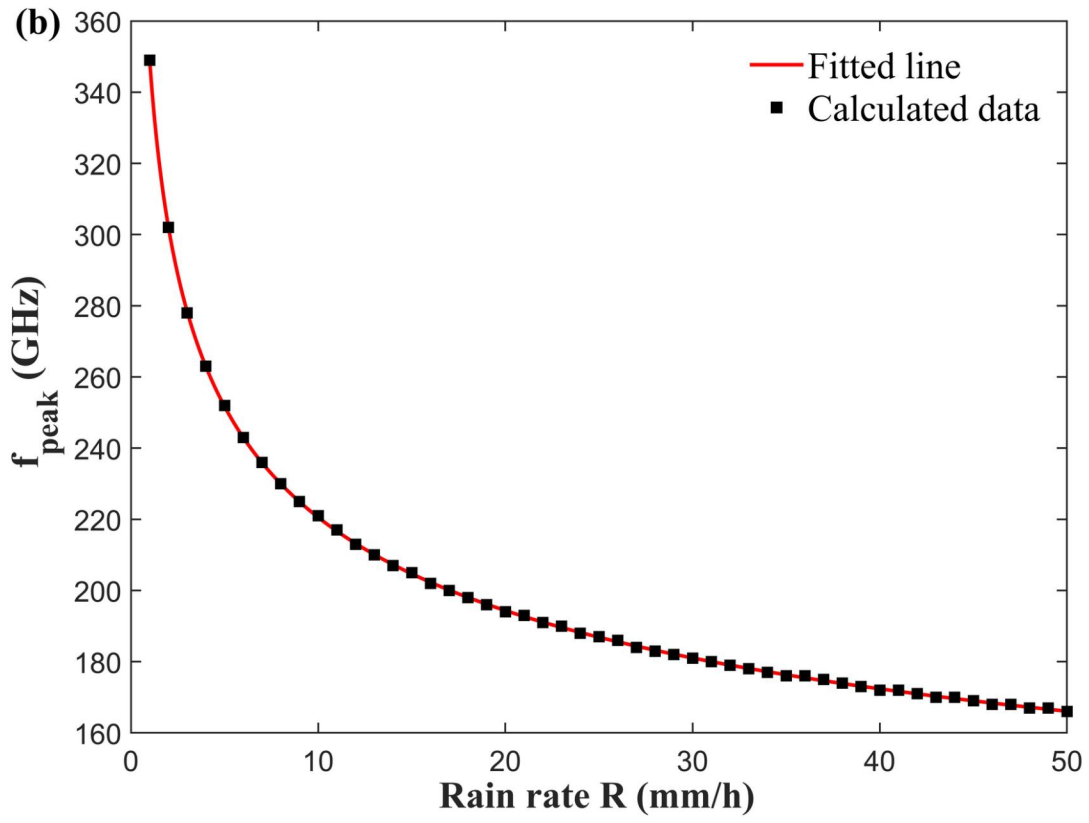

Figure 3. Relationships of the total-loss peak frequency $f_{peak}$ under the Best DSD model: (a) $f_{peak}$ versus the inverse characteristic scale $1/D_c$, with the calculated data and fitted line; (b) $f_{peak}$ versus rainfall rate R, with the calculated data and fitted curve.

## 4 Cross-Model Scale Interpretation

Figure 4(a) shows the total-loss peak frequency as a function of rainfall rate for all eight empirical DSD models at $T$ = 27°C. The curves do not coincide, but they share a consistent qualitative behavior. For every model, $f_{peak}$ decreases monotonically with increasing rainfall-rate $R$, and the rate of decrease gradually weakens in the high-rainfall regime. Downward peak migration is therefore a robust cross-model feature. The absolute peak level and the migration rate, however, remain DSD-dependent. This establishes that rainfall rate alone does not determine the peak position - the underlying droplet-size scale and DSD structure are equally essential. The ITU-R P.838-3 recommendation provides an independent anchor. The peak frequency implied by its published $k$ and $\alpha$ coefficients (see Section 1) decreases from 231 GHz at $R$ = 1 mm/h to 190 GHz at 10 mm/h and 149 GHz at 50 mm/h, and these results are overlaid with those shown in Fig. 4(a). It lies within the envelope spanned by the eight DSD models (approximately 156-484 GHz at $R = 1$ mm/h, narrowing to 85-202 GHz at 50 mm/h) and migrates more slowly than most individual models, consistent with the fixed representative drop spectra underlying the recommendation (ITU-R, 2003). The DSD-resolved curves can thus be read as the microphysical spread around the ITU-R average.

The M-P and J-W models provide an instructive limiting case. Their $f_{peak}$ curves in Fig. 4(a) overlap exactly. This follows directly from their parameterizations. Both models share the same slope parameter $\Lambda(R)$ and differ only by a constant amplitude factor in $N_0$ (see Table 2). An amplitude difference in $N_0$ scales the attenuation magnitude uniformly across frequency but does not alter the spectral shape. The peak location is therefore unchanged. This result corroborates the conclusion of Section 3 - $f_{peak}$ is governed by the DSD scale and shape, not by the overall number concentration.

To interpret the inter-model differences in quantitative terms, two representative droplet-size statistics are considered: the number-weighted mean diameter $\bar{D}$ (Wang et al., 2024) and the number median diameter $D_{50}(R)$ (Angulo-Martínez et al., 2018) are defined, respectively, as

$$\bar{D}(R) = \frac{\int_{D_{min}}^{D_{max}} DN(D;R)dD}{\int_{D_{min}}^{D_{max}} N(D;R)dD} \qquad (14)$$

$$\int_{D_{min}}^{D_{50}} N(D;R)dD = \frac{1}{2}\int_{D_{min}}^{D_{max}} N(D;R)dD \quad (15)$$

The rainfall-rate dependence of $\bar{D}$ and $D_{50}$ is shown in Figs. 4(b) and 4(c). For all empirical DSD models, both statistics increase monotonically with $R$, while $f_{peak}$ decreases. Within each model, peak-frequency migration is therefore consistently associated with growth of the representative droplet-size scale, and at any fixed rainfall rate, models with larger size statistics generally exhibit lower peak frequencies. The vertical separation among the $f_{peak}$-$R$ curves in Fig. 4(a) thus reflects inter-model differences in droplet-size scale rather than an isolated spectral effect. This within-model association is made explicit in Fig. 4(d). For the Best model, the total-loss $f_{peak}$ is well described by a linear function of $1/D_{50}$, mirroring the $1/D_c$ linearity established in Fig. 3(a), since any monotone size statistic of a fixed-shape DSD family tracks the characteristic scale $D_c$.

Across models, however, the two statistics are not equivalent. Pooling all eight models over $R = 1$-50 mm/h and regressing $f_{peak}$ against the inverse of each statistic with a single line yields $R^2 = 0.83$ for $\bar{D}$ and only 0.58 for $D_{50}$. The weakness of $D_{50}$ has a structural cause. For the Best model the number density diverges toward small drops, so the number median is pinned near the lower truncation - $D_{50} \approx 0.20$-0.23 mm almost independently of $R$, close to the analytic truncated-power-law value $2^{4/3}D_{min} \approx 0.25$ mm - and therefore reflects the choice of $D_{min}$ rather than a physical scale of the spectrum, visible in Fig. 4(d) as the narrow $1/D_{50}$ span (4.5-4.9 $mm^{-1}$) traversed over the full rainfall range. The clean linearity of Fig. 4(d) is thus a within-family property that does not transfer across models. Within individual DSD families, this size dependence is the same one represented in Section 3 by the model-internal scale $D_c$. Across families, $\bar{D}$ provides a directly computable, if imperfect, common descriptor for interpreting peak-frequency variation (Lee et al., 2004; Marshall & Palmer, 1948).

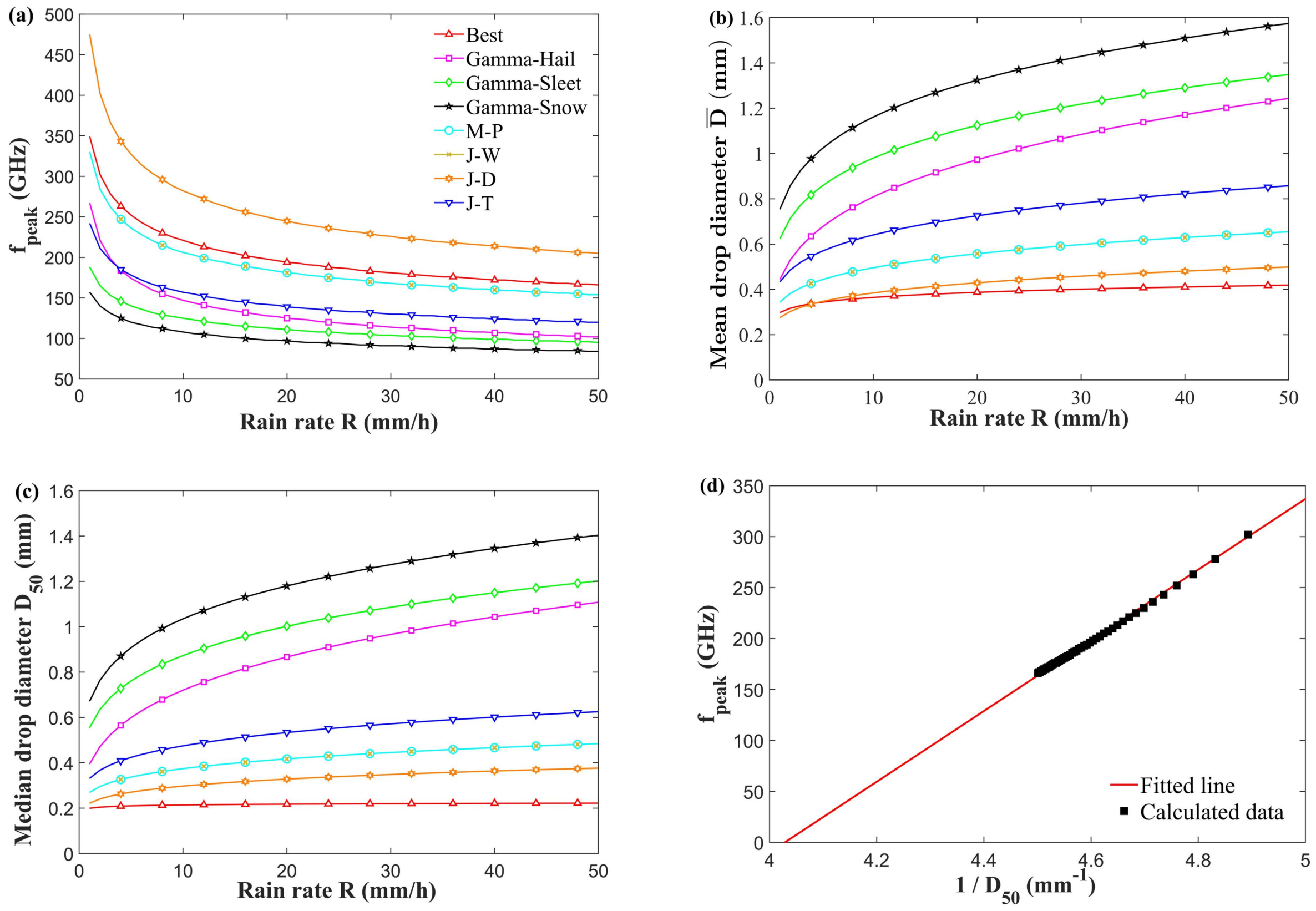

Figure 4. Rainfall-rate dependence of peak frequency and representative raindrop-size statistics for different empirical DSD models: (a) total-loss peak frequency $f_{peak}$ versus rainfall rate $R$; (b) number-weighted mean diameter $\overline{D}$ versus $R$; (c) number median diameter $D_{50}$ versus $R$; (d) relationship between $f_{peak}$ and the inverse number median diameter $1/D_{50}$, with the calculated data and fitted line.

## 5 Temperature Dependence

Sections 3 and 4 considered $T$=27°C in order to isolate the effect of rainfall-dependent DSD evolution on peak-frequency migration. In outdoor environments, however, rain events occur over a finite temperature range, and the temperature sensitivity of the peak location must therefore be assessed separately. At fixed rainfall rate and fixed DSD model, changing temperature does not modify the prescribed DSD itself. Its effect enters through the temperature dependence of the complex refractive index of liquid water, which in turn changes the Mie extinction response. This is fully consistent with the fixed-temperature dispersion result of Section 3. The peak is insensitive to the gradual variation of $m$ with frequency across the peak region at fixed $T$, but it responds to the global change of $m$ with temperature, which is driven by the primary Debye relaxation of liquid water - the relaxation frequency rises from 8.9 GHz at 0°C to 19.3 GHz at 25°C (Liebe et al., 1991) - and rescales the absorptive response across the entire band, shifting the family constants $x_j^*$. To quantify this effect, the $f_{peak}$-$R$ relation is recalculated at $T$= 1°C, 10°C, 20°C, and 30°C.

Figure 5 shows the representative result for the Best DSD model, and qualitatively similar behavior is obtained for the other empirical DSD models. The temperature effect is rainfall-dependent. At high rainfall rates the shift is clear and upward. At $R = 50$ mm/h, $f_{peak}$ rises from 149.2 GHz at 1°C to 165.8 GHz at 30°C, approximately +0.57 GHz/°C. At intermediate rainfall rates the sensitivity weakens markedly ($|\partial f_{peak}/\partial T| \leq 0.1$ GHz/°C at $R = 10$ mm/h, with non-monotonic behavior), and at the lowest rainfall rates - where the spectral maximum is broad - the computed trend is slightly reversed (377.6 GHz at 1°C versus 353.9 GHz at 30°C at $R = 1$ mm/h). Temperature therefore acts as a secondary, rainfall-dependent correction that is most pronounced at high rainfall rates, and it does not change the basic migration law. At all temperatures considered, $f_{peak}$ decreases monotonically with increasing rainfall rate, and the curves retain nearly the same shape over the full rainfall range.

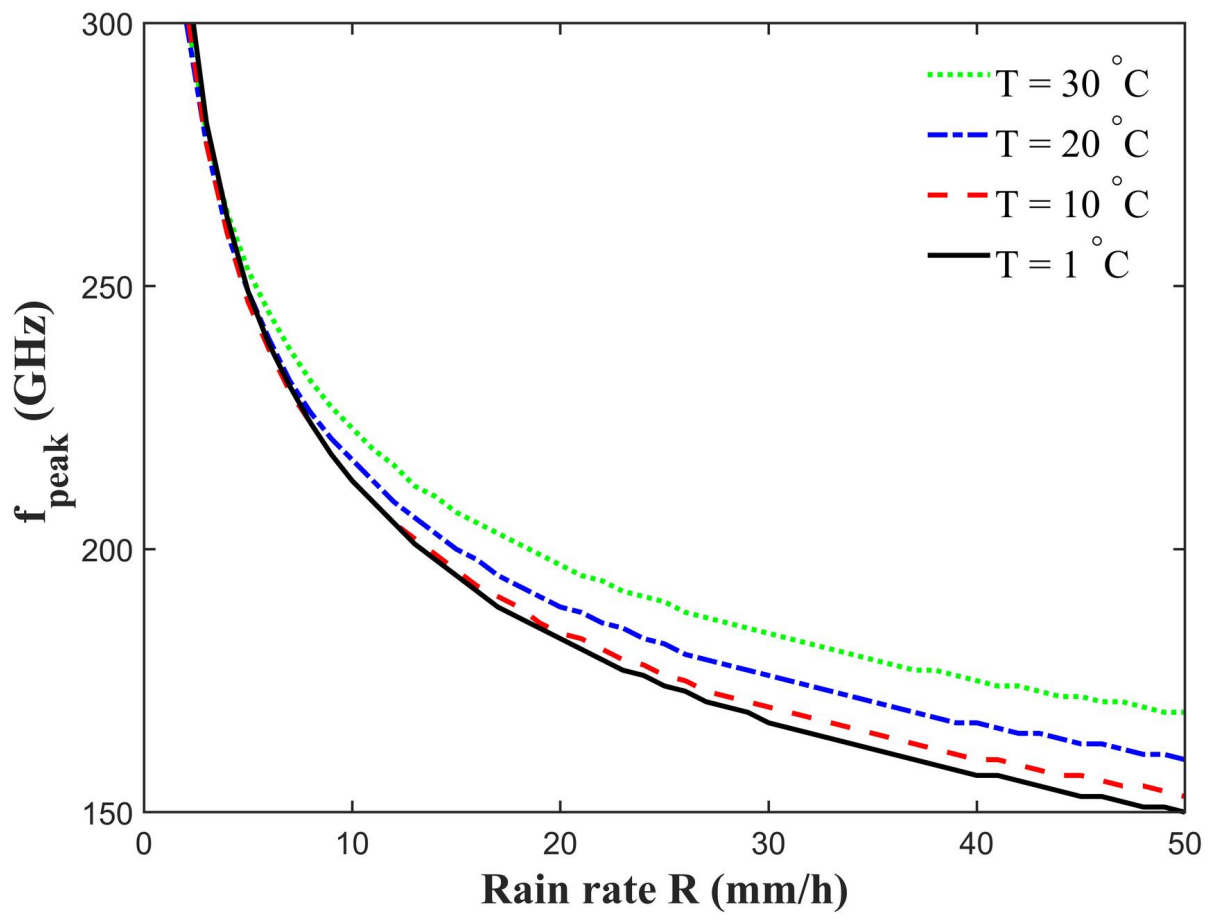

Figure 5. Relationship between the total-loss peak frequency $f_{peak}$ and rainfall rate $R$ at different temperatures under the Best DSD model.

Table 5. Best-fit APL parameters for the Best DSD model at five temperatures, extracted with the parabolic peak refinement of Section 2.

| $T$ (°C) | $f_\infty$ (GHz) | $B$ | $p$ | RMSE | $R^2$ |
|---|---|---|---|---|---|
| 1 | -15.32 | 393.53 | 0.224 | 0.257 | 0.99997 |
| 10 | 29.42 | 346.59 | 0.271 | 0.624 | 0.99981 |
| 20 | 73.62 | 292.93 | 0.323 | 0.244 | 0.99997 |
| 27 | 87.18 | 269.82 | 0.324 | 0.191 | 0.99997 |
| 30 | 89.67 | 263.53 | 0.316 | 0.263 | 0.99995 |

From a modeling perspective, this means that the fixed-temperature APL-type description developed above remains structurally valid under different thermal conditions, with temperature entering through its fitted parameters. Table 5 lists the APL parameters refitted at each temperature for the Best model. $f_\infty$ increases and $B$ decreases systematically with temperature, while the migration law itself is preserved. In other words, temperature should be incorporated as a calibration dimension of the peak-frequency model, rather than as a factor requiring a different migration law. In THz communications, this means that temperature does not require a new rain-attenuation peak model, but it does affect where the most strongly impaired spectral region appears - increasingly so at high rainfall rates. As a result, weather-aware frequency planning (Liu & Huang, 2025) and adaptive subband selection (Kumar, 2023) should account for temperature through parameter recalibration, especially in wideband outdoor THz links.

## 6 Conclusions

Rain attenuation in the THz band is inherently frequency selective, so characterizing only the attenuation magnitude at a few discrete frequencies is insufficient for wideband link analysis. In this article, the spectral peak frequency of rain attenuation was investigated as a compact channel descriptor using Mie-theory-based calculations under one laboratory Gaussian DSD and eight outdoor empirical DSD models. For the laboratory Gaussian case, where the DSD shape is fixed and only the total number concentration changes with rainfall rate, the attenuation spectrum preserves a stable peak location while its magnitude increases. In contrast, for all outdoor empirical DSD models, the total-loss peak frequency decreases monotonically with increasing rainfall rate - natural rainfall reshapes the spectral structure of rain-induced loss, not merely its level.

Further analysis showed that this migration is governed primarily by the rainfall-dependent DSD characteristic scale and follows a quasi-exact inverse scaling with that scale, whereas the explicit frequency dispersion of the water refractive index produces no measurable shift in $f_{\mathrm{peak}}$ at fixed temperature under the present computational settings. The peak-frequency evolution is well captured by an asymptotic power-law expression, preferred over Hill and pure power-law alternatives by AIC and BIC. Across DSD families, the number-weighted mean diameter $\bar{D}$ tracks the peak frequency more reliably than the number median diameter $D_{50}$, which is dominated by the lower truncation of the drop spectrum, although residual family-shape dependence persists in either case. Temperature acts as a secondary, rainfall-dependent correction - most pronounced at high rainfall rates - that recalibrates the APL parameters without altering the migration law, and the peak frequency implied by the ITU-R P.838-3 coefficients falls within the DSD-model envelope while migrating more slowly, anchoring the present results to the standard engineering model. Because gaseous absorption defines the usable transmission windows above 100 GHz, a natural next step is to rank candidate windows under combined rain and gaseous loss, with the rain contribution summarized by $f_{\mathrm{peak}}(R)$ and its prominence width, supporting weather-aware band selection, adaptive subband allocation, and robust link design in future wideband THz communication systems.

## Appendix. Supplementary results

Figure A1 presents attenuation characteristics for the M-P and Gamma-Hail models, supplementing the Best-model results in the main text. In both cases, the total-loss spectra in Figs. A1(a) and (c) retain a clear single-peak structure over $R$ = 1-50 mm/h, with the peak frequency shifting progressively toward lower frequencies as rainfall rate increases. The component-wise results in Figs. A1(b) and (d) confirm that the scattering peak undergoes the strongest downward migration, the absorption peak varies more gradually, and the total-loss peak remains intermediate. The absolute peak-frequency levels differ between the two models, reflecting their distinct droplet-size scales, but the overall migration trend is shared. These results confirm that downward peak migration is a robust feature of outdoor empirical DSD conditions, independent of the specific model family.

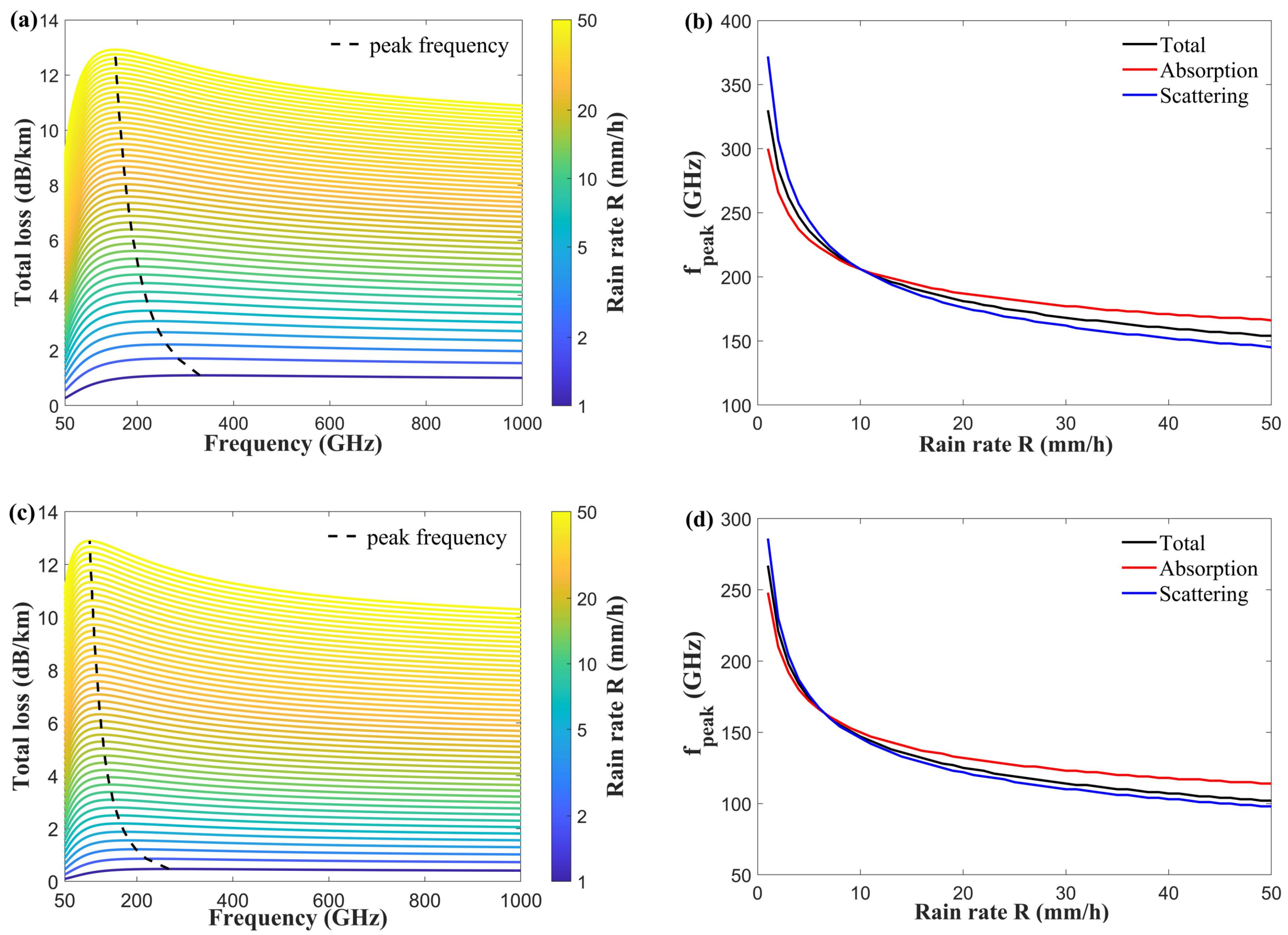

Figure A1. Rain-attenuation characteristics under the M–P and Gamma-Hail DSD models at $T$ = 27 °C, supplementing the Best-model results in the main text: (a) total-loss rain-attenuation spectra over 50–1000 GHz for rainfall rates of 1–50 mm/h under the M–P model, with the migrating peak frequency indicated; (b) peak frequencies of the total-loss, absorption, and scattering components as functions of rainfall rate $R$ under the M–P model; (c) total-loss rain-attenuation spectra under the Gamma-Hail model; (d) peak frequencies of the total-loss, absorption, and scattering components as functions of rainfall rate R under the Gamma-Hail model.

Figure A2 compares the total-loss $f_{peak}$–R curves calculated using the full frequency-dependent refractive index $m(f, T)$ and a fixed refractive index $m(f_{ref}, T)$ for the Best DSD model at $T$ = 27°C. The two curves are pointwise identical to the reported numerical precision over the full rainfall-rate range. The same result was confirmed for all eight empirical DSD models, all three reference frequencies ($f_{ref}$ = 140, 220, and 340 GHz), and all three loss components (total loss, absorption, and scattering). Thus, under the present computational settings, the explicit frequency dispersion of the water refractive index does not measurably affect the location of $f_{peak}$ at fixed temperature. This result supports the use of a fixed representative refractive index in the peak-location scaling analysis of Section 3; it does not imply that dielectric dispersion has no effect on the attenuation magnitude or on other spectral characteristics.

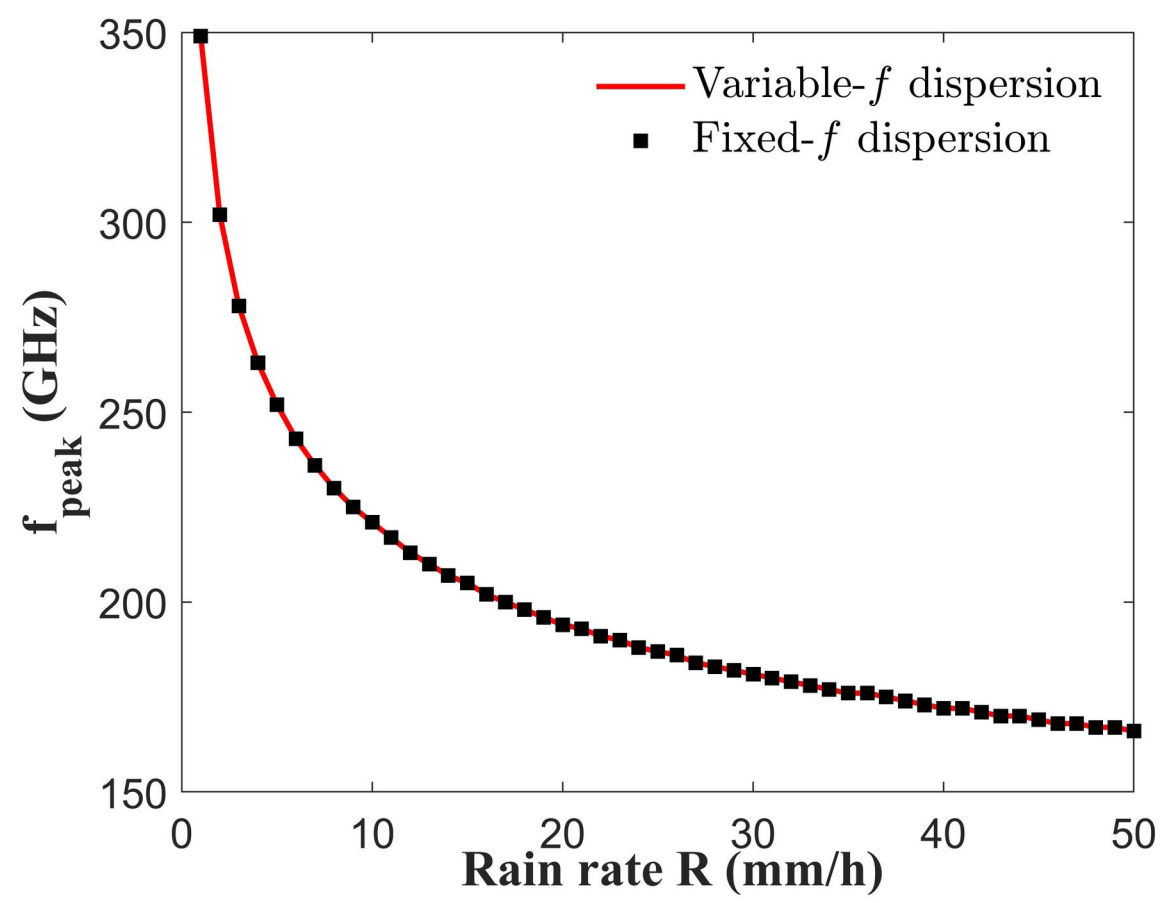

Figure A2. Comparison of the extracted peak frequency under the variable-frequency and fixed-frequency dispersion treatments at $T = 27$ °C.

## Acknowledgments

This work was supported in part by the National Natural Science Foundation of China under Grant 62471033, the Natural Science Foundation of Hebei Province under Grant F2026105021, and the Special Program Project for Original Basic Interdisciplinary Innovation under the Science and Technology Innovation Plan of Beijing Institute of Technology under Grant 2025CX11010.

## Open Research

Data underlying the results presented in this paper can be obtained from the corresponding author - Jianjun Ma upon reasonable request.

## Conflict of Interest Disclosure

The authors declare there are no conflicts of interest for this manuscript.